\documentclass[twocolumn,amssymb,amsmath, %otv
tightenlines,showpacs,floatfix,
superscriptaddress,aps,prl]{revtex4-1}  %oc

\usepackage{graphicx} 
\usepackage{dcolumn}% Align table columns on decimal point
\usepackage{bm}% bold math

\begin{document}

\title{Annealing of supersolidity in plastically deformed solid $^4$He} %otv
\author{Debabrata Sinha} %otv scheme changed
\email{debabratas@tifrh.res.in}
\author{Surajit Sengupta}
\email{surajit@tifrh.res.in}
\altaffiliation{Also at Centre for Advanced Materials, Indian Association for Cultivation of Science, Jadavpur, Kolkata 700032, India}
\affiliation{TIFR Centre for Interdisiplinary Sciences, 21 Brundavan Colony, Narsingi, Hyderabad 500075, India}
\author{Chandan Dasgupta}
\email{cdgupta@physics.iisc.ernet.in}
\altaffiliation{Also at Jawaharlal Nehru Centre for Advanced Research, Bangalore 560064,
India}
\affiliation{Centre for Condensed Matter Theory, Department of Physics, Indian Institute of Science, Bangalore 560012, India}
\author{Oriol T. Valls}
\email{otvalls@umn.edu}
\altaffiliation{Also at Minnesota Supercomputer Institute, University of Minnesota,
Minneapolis, Minnesota 55455}
\affiliation{School of Physics and Astronomy, University of Minnesota, 
Minneapolis, Minnesota 55455}
\date{\today}
\begin{abstract} %oc many rewrites
We present  a numerical study of a continuum  plasticity field coupled to a Ginzburg-Landau model for superfluidity. The results suggest that a supersolid fraction may appear as a long-lived transient during the time evolution of the plasticity field at higher temperatures where both dislocation climb and glide are allowed. Supersolidity, however, vanishes with annealing. As the temperature is decreased, dislocation climb is arrested and any residual supersolidity due to incomplete annealing remains frozen. Our results provide a resolution of many perplexing issues concerning a variety of experiments on solid $^4$He.  
\end{abstract}

\maketitle
% introduction _ general

%Many years after the original proposal by Chester\cite{chester} and Legget\cite{Legget} that supersolidity could exist in bosonic
%solids, 
The first reports of non-classical rotational inertia (NCRI) in solid ${\rm ^4He}$, \cite{sup-kim-1,sup-kim} interpreted as evidence
for a ``supersolid'' phase,  triggered an avalanche of additional work \cite{sup-others} which largely confirmed the existence
of an anomalous NCRI signal in the form of a  sharp drop in the period  of a torsional oscillator (TO) filled with solid $^4$He. 
%ovcd Minor edits and rearranging citations below
%The He consisted either of crystallites frozen within porous Vycor glass\cite{sup-kim-1} or of 
%bulk polycrystal\cite{sup-kim, sup-others} below 200 mK. 
However, the nature and origin of the TO anomalies have been very much in dispute, especially after the experimental observation \cite{elastic} of
an increase in the shear modulus of solid $^4$He at the onset of the NCRI. 
The period anomalies can be due to the appearance of a new supersolid phase  (producing an  NCRI fraction, NCRIF, 
%in the solid  $^4$He, 
which decouples from the TO), to changes in the elastic properties, or to a combination of both.  It is not at all easy to disentangle these two effects.
%as experimental evidence showing 
%that\cite{elastic} they are related was found, and  it became evident\cite{Dorsey,Toner,Dasbiswas} that elastic and superfluid 
%properties were very likely coupled. 
At one extreme it has been suggested that the phenomena involved are largely or exclusively elastic anomalies \cite{no-ncrif}. 
On the other hand, experimental evidence for the mass flow equivalent of a fountain effect \cite{mass-flux}, which is in principle the gold 
standard for the existence of a superfluid field, indicates the presence of superfluidity. Recent experiments \cite{rotation} on the effects of
dc rotation on the observed NCRIF also suggest the occurrence of superfluidity in solid  $^4$He.
%It has not yet been
%possible to completely rule out that the observed fountain
%effect is not due to the presence of liquid material flowing between
%microcrystals. 

In our view, the above alternatives may be a false dichotomy. The apparent close connection between the elastic properties of the solid 
and those of the putative superfluid field may be understood %ovcd rationalized 
if one assumes that extended  crystal defects, such as dislocations and
grain boundaries, play a crucial role in the observed phenomena. 
%While experiments\cite{rule-out} have
%ruled out the possibility of local melting either at the 
%surface or within grain boundaries, 
A prominent role for crystal defects \cite{sample,anneal} is indicated by several experimental and theoretical results. 
The NCRIF reported varies dramatically from sample to sample \cite{sample}, and can decrease substantially on annealing\cite{anneal}.  
Also, quantum Monte Carlo \cite{gb-super-qmc,core} calculations have shown that superfluidity can occur along dislocation cores and near grain boundaries.
As posited in Ref.~\onlinecite{feff}, the apparent contradictions may be explained by assuming that the motion of dislocations, known to
affect the elastic properties of the solid, also has a strong effect on the occurrence of superfluidity along their cores.
The reported \cite{mass-flux}  mass flow through solid $^4$He has been attributed to 
quantum superclimb arising from the flow of atoms though superfluid dislocation cores \cite{superclimb}.

In this Letter we present the results of a calculation that points to a reasonable resolution of these dilemmas by modeling the 
dynamics of a large number of dislocations using a recent formulation of 
continuum plasticity theory \cite{acharya,Lim-1,Lim-2,Lim-GB, Lim-CW, Lim-thesis} coupled to a complex scalar field $\psi$ \cite{Dorsey,Toner, Dasbiswas}
that describes superfluid order. Our argument is %ovcd essentially 
as follows: It is known \cite{Lim-2} that an initial smooth distribution of 
dislocations spontaneously coarsens into defect-free regions interspersed with shock-like structures of high dislocation density or ``cell walls''. 
Internal stresses at such cell walls, which may be large initially, eventually anneal out producing stress free-grain boundaries 
(representing discontinuities only in crystal orientation) at late times \cite{Lim-GB,Lim-CW}. If $\psi$ couples only to volumetric stress $\sigma_{ii}$ (the
trace of the stress tensor), %ovcd
as in \cite{Toner}, transient superfluidity may exist at cell walls with large $\sigma_{ii}$ as long as dislocation climb is allowed. 
Once cell walls evolve into symmetric grain boundaries, this superfluidity vanishes, %ovcd
in agreement with quantum Monte Carlo 
results \cite{gb-super-qmc}. If climb is arrested, the plastic current is volume preserving \cite{Lim-GB, Lim-CW} and $\psi$, if initially absent, 
cannot form. Similarly, annealing of an initially non-zero $\psi$ is severely constrained without dislocation climb. This suggests the following 
scenario for the occurrence of NCRI: %ovcd
At high temperatures, dislocation climb, ensured through mass transport through dislocation 
cores \cite{superclimb}, 
results in the formation of symmetric grain boundaries and vanishing of the 
supersolid fraction at long times. As the temperature is
reduced and/or pinning of dislocations by impurities becomes effective, %ovcd
climb is suppressed and the resulting long-lived residual 
supersolidity at cell walls contribute to the NCRI.

The Ginzburg-Landau free-energy for the complex superfluid order-parameter $\psi({\bf r})$ coupled to the elastic displacement ${\bf u}_i$ of the solid is \cite{Dorsey,Toner}, 
\begin{eqnarray}
\mathcal{H} = \int d {\bf r} \left[ c_0\, \vert \nabla \psi \vert^2  + \frac{a_{0}}{2}|\psi|^{2}+\frac{d_{0}}{4}|\psi|^{4} \right] + \mathcal{H}_{int}
\label{LG}
\end{eqnarray}
The parameters $c_{0}$,$a_{0}$ and $d_{0}$ all are temperature dependent and the interaction energy is 
\begin{eqnarray}
\mathcal{H}_{int}=\frac{1}{2}g_{ij}\int d {\bf r} u_{ij}|\psi|^{2}, %ovcd
\label{int}
\end{eqnarray}
where $u_{ij}=\frac{1}{2} (\partial_{i} u_{j}+\partial_{j} u_{i})$ and $g_{ij}$ couple the elastic degrees of freedom to the magnitude of $\psi$. To the extent that NCRI is caused by supersolidity of $^4$He,  the norm $|\psi|^2$ of the order parameter is expected to be related to the NCRIF measured in experiments. %Hereafter we simply define NCRIF to be so. 

As a consequence of the coupling in (\ref{int}), the solid experiences an ``external'' stess due to $\psi$, 
\begin{eqnarray}
\sigma^{ex}_{ij}={\frac{\delta \mathcal{H}_{int}} %otv over 
{\delta u_{ij}}}=\frac{1}{2}g_{ij}\vert \psi\vert ^{2}
\label{super-sig}
\end{eqnarray}
Dislocations %otv
within the solid respond to both $\sigma^{ex}_{ij}$ and to the %otv
stress due to the presence of other dislocations. To address the problem of defect dynamics in a solid, we follow the recent work of Acharya \cite{acharya} and that of Limkumnerd and Sethna \cite{Lim-1,Lim-2,Lim-GB,Lim-CW,Lim-thesis}. It is convenient to introduce the plastic deformation field \cite{acharya} $\beta_{ij}^p$, such that the total deformation gradient $\partial_j u_i = \beta_{ij}^e + \beta_{ij}^p$ is a sum of elastic and plastic parts\cite{Lim-2}. The dislocation density tensor is $\rho_{ij} ({\bf r}) =   \sum_{\gamma} t_i^{\gamma} b_j^{\gamma} \delta({\bf r} - {\bf r}^{\gamma}) = -\varepsilon_{ilm} \partial \beta^p_{mj}/\partial r_l$, where the vectors ${\bf t}^{\gamma}$ and ${\bf b}^{\gamma}$ are the tangent and Burgers vectors respectively of a dislocation line at ${\bf r}^{\gamma}$ and $\varepsilon_{ilm}$ is the antisymmetric tensor. Dynamical equations for $\beta_{ij}^p$ can now be derived after writing the current  $\dot \beta_{ij}^p = \sum_{\gamma} J_{ij}^{\gamma} = \sum_{\gamma} {\varepsilon}_{ilm} t_l^{\gamma} b_j^{\gamma} v_m^{\gamma} \delta({\bf r} - {\bf r}^{\gamma}) $ %ovcd alpha missing in v?
as a sum over single dislocation contributions, with the velocity of 
a single dislocation line $v_i %ovcd PK not introduced yet = D f^{PK}_i 
= - D {\varepsilon}_{ijk} t_j b_l \sigma_{kl}$ %ovcd alphas missing
proportional to the {\em total}
 stress $\sigma_{ij}= \sigma^d_{ij} + \sigma^{ex}_{ij}$ consisting of separate contributions from other dislocations ($\sigma^d_{ij}$) and from $\psi$; $D^{-1}$ is a material dependent time scale over which plasticity anneals. Apart from being driven in the direction of  ${\bf b}$ by the local stress (glide), dislocations may also {\em climb} i.e. move in the perpendicular direction in response to the local flux of point defects. %ovcd rewrite below
Total volume is preserved by glide, but inclusion of climb  removes this constraint. The difference between glide and climb motion was incorporated phenomenologically in Ref.~\onlinecite{Lim-2} by writing the total flux as a sum of two terms,     
\begin{eqnarray}
J_{ij}^{\gamma} &=& D [\epsilon_{ilm}t_{l}^{\gamma}b_{j}^{\gamma}\epsilon_{mpq}\sigma_{pr}t_{q}^{\gamma}b_{r}^{\gamma}\delta({\bf r} - {\bf r}^{\gamma}) \nonumber\\
& & -\frac{\lambda}{3}\delta_{ij}\epsilon_{klm}t_{l}^{\gamma}b_{k}^{\gamma}\epsilon_{mpq}\sigma_{pr}t_{q}^{\gamma}b_{r}^{\gamma}\delta ({\bf r} - {\bf r}^{\gamma})]
\label{current}
\end{eqnarray}
For $\lambda=1$, $J_{ij}$ is traceless i.e only glide motion is possible and for $\lambda=0$ glide and climb are equally probable \cite{acharya}. Next, %ovcd
coarse graining the current over all dislocations, within a mean-field approximation, 
one obtains
\begin{eqnarray}
\partial_{t}\beta^{p}_{ij} & = & \frac{D}{2} \Big[(\sigma_{ic}\rho_{ac}-\sigma_{ac}\rho_{ic})\rho_{aj} - \nonumber \\ 
& & \frac{\lambda}{3}\delta_{ij}(\sigma_{kc}\rho_{ac}-\sigma_{ac}\rho_{kc})\rho_{ak})\Big]
\label{plas-dyn}
\end{eqnarray}
which is identical to that obtained in \cite{Lim-2} except for a redefinition of $\sigma_{ij}$. The stress due to dislocations is $\sigma^{d}_{ij}=-\bar{C}_{ijkm}\beta^{p}_{km}$%otv  
%\begin{eqnarray}
%\sigma^{d}_{ij}=-\bar{C}_{ijkm}\beta^{p}_{km}, %otv
%\label{hooks}
%\end{eqnarray}
where, for an isotropic solid, 
\begin{eqnarray}
\bar{C}_{ijkm} & = & \mu (\delta_{ik} \delta_{jm}+ \delta_{im} \delta_{jk}+\frac{2\nu}{1-\nu} \delta_{ij} \delta_{km}) \nonumber \\
\end{eqnarray}
and $\mu$ and $\nu$ are the shear modulus and Poisson's ratio respectively. In Ref.~\onlinecite{Lim-2} it was found that an %ovcd
equation similar to (\ref{plas-dyn}) %ovcd
spontaneously leads to dislocation pile-ups with associated stress jumps \cite{Lim-CW}, known as cell walls, %ovcd
at finite time. At later times, %ovcd
cell walls %ovcd (see 2 lines below)
evolved %ovcd
to symmetric stress-free grain boundaries \cite{Lim-GB} 
by attracting more and more dislocations to themselves, %ovcd
as in the %ovcd
formation of finite time shocks in Burgers turbulence \cite{burgulence}. %ovcd

In principle, Eq.~(\ref{plas-dyn}) %ovcd
together with a dynamical equation for $\psi$ such as the time dependent Ginzburg-Landau (TDGL) equation $\partial_{t}\psi=-\Gamma \delta\mathcal{H}/\delta \psi$,
%\begin{eqnarray}
%\partial_{t}\psi={\delta \mathcal{H}\over \delta \psi}=-c \partial_{z}^{2}\psi+ a\psi+d |\psi|^{2}\psi
%\label{order}
%\end{eqnarray}
is sufficient to describe the dynamics of a supersolid %ovcd
with plastic deformation. The parameter $\Gamma$ sets the time scale for the evolution of $\psi$. %ovcd
To make explicit calculations, we now introduce two simplifications. Firstly, since stress $\sigma_{ij} = \partial {\mathcal H}/\partial u_{ij}$ relaxes much faster than the dislocation configuration, we may assume that for all times, the divergence of the total stress vanishes, i.e,  $\partial_{j}\sigma_{ij}=0$ and the solid is in {\em mechanical equilibrium}. Secondly, we report calculations only for situations where all the fields, $\beta^p_{ij}, \psi$ etc. are functions only of {\em one} dimension, $z$ describing {\em flat} cell walls and grain boundaries \cite{Lim-CW,Lim-GB}. %otv that 
Using $\sigma^{ex}_{ij}$ and $\sigma^d_{ij}$ in the mechanical equilibrium condition, we obtain, in one dimension,
\begin{eqnarray}
%\vec{\nabla}.\vec{u}
\frac{\partial u_z}{\partial z} & = &-\frac{\nu}{1-\nu}\beta^{p}_{kk} - \frac{(1-2\nu)}{4\mu(1-\nu)}g|\psi|^{2},
\label{diver}
\end{eqnarray}
where $\beta^p_{kk} \equiv \beta^p_{11}+\beta^p_{22}$. %ovcd in addition, we have %otv
We have simplified ${\mathcal H}_{int}$ by %otv
assuming that $\psi$ couples only to the volumetric stress $\sigma_{ii} = [2 \mu (1+\nu)/3 (1- 2 \nu)] \partial_z u_z$. %ovcd
A similar assumption was  used by Toner \cite{Toner} %otv
with the sign of $g$ chosen such that compressive stresses lead to superfluidity. 
The Hamiltonian is now given by,%otv
\begin{eqnarray}
\mathcal{H} & = & \mathcal{H}_{LG}+\mathcal{H}_{int}=\int dz\,\, c_0 \left(\frac{\partial \psi}{\partial z}\right)^{2}+\frac{a}{2}|\psi|^{2}+\frac{d}{4}|\psi|^{4}, \nonumber \\
\label{total}
\end{eqnarray}
with renormalized parameters as a consequence of Eq.~(\ref{diver}) viz., $a = a_{0}-g(\nu/1-\nu)\beta^{p}_{kk}$ and $d = d_{0}-g^{2}[(1-2\nu)/\mu(1-\nu)]$.
%\begin{eqnarray}
%a &=& a_{0}-g\frac{\nu}{1-\nu}\beta^{p}_{kk}\\
%d &=& d_{0}-g^{2}\frac{2(1+\nu)}{3K(1-\nu)}
%\end{eqnarray}
The constants $a_0$ and $d_0$ are both $>0$, so that the equilibrium %ovcd
bulk solid does not show superfluidity at any temperature. The time evolution equation for $\beta^{p}_{ij}$ ($i,j = 1,2$)  and $\beta^p_{33}$ including superfluidity is given by,  
\begin{subequations}
\label{supersolid} 
\begin{align}%ovcd
-\partial_{t}\beta^{p}_{ij} & =  \frac{1}{2}(\partial_{z}\mathcal{E})\partial_{z}(\beta^{p}_{ij}-\frac{\lambda}{3}\beta^{p}_{kk}\delta_{ij}) -\frac{1}{4}g|\psi|^{2}\partial_{z}\beta^{p}_{ij}\partial_{z}\beta^{p}_{kk} \nonumber \\ 
&+ \frac{\lambda}{12}g|\psi|^{2}(\partial_{z}\beta^{p}_{kk})^{2}\delta_{ij}, \\
-\partial_t\beta^p_{33} & =  - \frac{\lambda}{6}\partial_{z}\mathcal{E}\partial_{z}\beta^{p}_{kk} + \frac{\lambda}{12} g|\psi|^{2}(\partial_{z}\beta^{p}_{kk})^{2}
%\label{supersolid}
\end{align}
\end{subequations}
with $\mathcal{E}=-\sigma^{d}_{ij}\beta^{p}_{ij}$ being %otv
the elastic energy density from dislocations.  %SS 
%We have now set $D=1$, which fixes our unit of time. %oc %o
For %ovcd
$\lambda = 0$, $\beta^p_{33}$ does not evolve and for $\lambda = 1$, the evolution is constrained such that the trace $\beta^p_{11} + \beta^p_{22} + \beta^p_{33}$ vanishes at all times. We choose $c_0$ and $D^{-1}$ as our units for length and time respectively, the energy scale is set by $k_B T$.
%\begin{eqnarray}
%\mathcal{E} & = &\mu [\frac{1}{2}(\beta^{p}_{12}+\beta^{p}_{21})^2+({\beta^{p}_{11}}^2+{\beta^{p}_{22}}^2)+\frac{\nu}{1-\nu}(\beta^{p}_{11}+\beta^{p}_{22})^2]
%\label{energy}
%\end{eqnarray}

We shall first consider the case with $\lambda=0$.  The components of $\beta^{p}_{ij}$ (see Eq.~(\ref{supersolid})) are all coupled through 
the Peach-K\"{o}hler force density (PKF) %ovcd define %otv
$\mathcal{F}^{PK}=\partial \mathcal{E}/\partial z$. %ovcd the PKF density. 
It is instructive, therefore, to obtain the time evolution of $\mathcal{F}^{PK}$. For $\lambda=0$ we obtain from Eq.~(\ref{supersolid}). %otv
\begin{equation} %otv
\partial_{t}\mathcal{F}^{PK}+\mathcal{F}^{PK}\partial_{z}\mathcal{F}^{PK}-
\frac{1}{4}g\partial_{z}(|\psi|^{2}\mathcal{F}^{PK}\partial_{z}\beta^{p}_{kk})
=0. %otv
\label{pk force}
\end{equation} %otv
For $g=0$  Eq.~(\ref{pk force}) %otv
becomes the Burgers equation \cite{burgulence}. 
%which develops finite time discontinuities.\cite{Lim-1,Lim-2} %otv
%A discontinuty %otv
%is the signature of cell wall formation\cite{Lim-GB,Lim-CW} where dislocations aggregate to form wall-like structures with sharp jumps in some stress components. 
\begin{figure}
\rotatebox{0}{\includegraphics[width=3.5in]{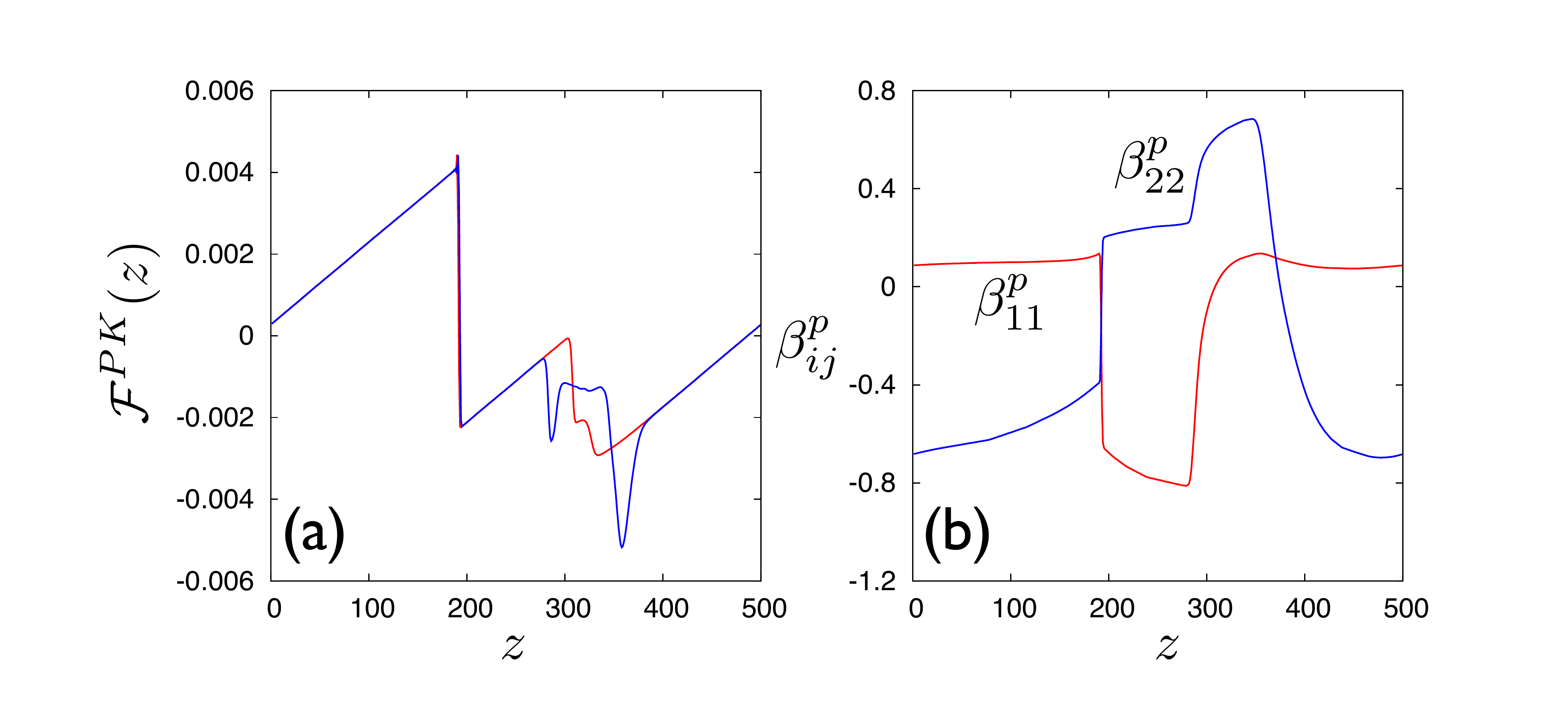}}
\caption{(Color-online) (a) Peach-K\"ohler force density ${\mathcal F}^{PK}$ (blue/dark gray curve) in the presence of supersolidity for $\lambda=0$ 
at reduced time $t= 50$ %oc 500\Delta t$
%ovcd said in text A fourth order Runge-Kutta scheme is used to 
as obtained from numerical integration of 
Eqs.~(\ref{supersolid}) and (\ref{pk force}). The red/light gray curve shows ${\mathcal F}^{PK}$ in the absence of $\psi$ (i.e. $g=0$) for the same initial conditions for comparison. (b) %ovcd 
plasticity tensor components %otv %CD inserted space between the beta's
$\beta^{p}_{11}$, $\beta^{p}_{22}$ from the same calculation as in (a).} %ovcd
\label{fig:f-and-beta}
\end{figure}

We have solved the coupled Eqs.~(\ref{supersolid}) and (\ref{pk force}) %otv
together with the TDGL equation for $\psi$, numerically using an accurate fourth order Runge-Kutta scheme \cite{NR} with time step 
$\Delta t = 0.05$ %oc D^{-1}$, 
 %ovcd
and spatial discretization $\Delta z = 1$. The initial input for $\beta^{p}_{ij}$ is random, drawn from a Gaussian distribution with mean zero and width $=0.15$. The equations are regularized  by adding a diffusive term $\alpha\, \partial_z^2 \beta^p_{ij}$ %ovcd? $\alpha \nabla^{2}\beta^{p}_{ij}$ 
with a small initial value of 
$\alpha$ %ovcd
which is subsequently reduced further at later times. The rest of the parameters  $\mu = 1$, $\nu = .49$, $a_0 = 0.01$, $g = .5$ and $d = 1$ are chosen to represent a generic solid above the bulk superfluid transition. Finally, for the dynamical parameters we use $\Gamma = D$ %oc
which represents a scenario where $|\psi|^2$ relaxes together with the plasticity and the associated stress. 
%We comment further on these parameter choices later. 

In Fig. \ref{fig:f-and-beta}(a) and (b)  we plot,
respectively, %otv
the PKF and the two diagonal components of $\beta^{p}_{ij}$. Both the PKF and the $\beta^p_{ij}$ have %otv
two discontinuities at $z =180$ and $\approx 300$ which are inherited from the Burgers-like terms in Eq.~(\ref{supersolid}) modified by the presence of $\psi$. %otv
%indeed, all components of $\beta^{p}_{ij}$ are discontinuous. 
Since the dislocation density $\rho_{ij}$ is given by $z$ derivatives of 
$\beta_{ij}^p$, %ovcd defined below 3, ref not needed \cite{note}, %ovcd
these regions of $\beta^p_{ij}$ are also regions of large dislocation content signifying the presence of dislocation pile-ups or cell walls \cite{Lim-CW}. 
%ovcd Interestingly, 
While the first of these cell walls is, coincidentally, almost free of compressive stresses, the second has a prominent stress jump which reduces $a$ in Eqn.(\ref{total}) locally. The location and nature of the discontinuities depend, of course, %ovcd
on the realization of the random initial condition used. 
%otv paragraph berak added
 
\begin{figure}
\rotatebox{0}{\includegraphics[width=3.5in]{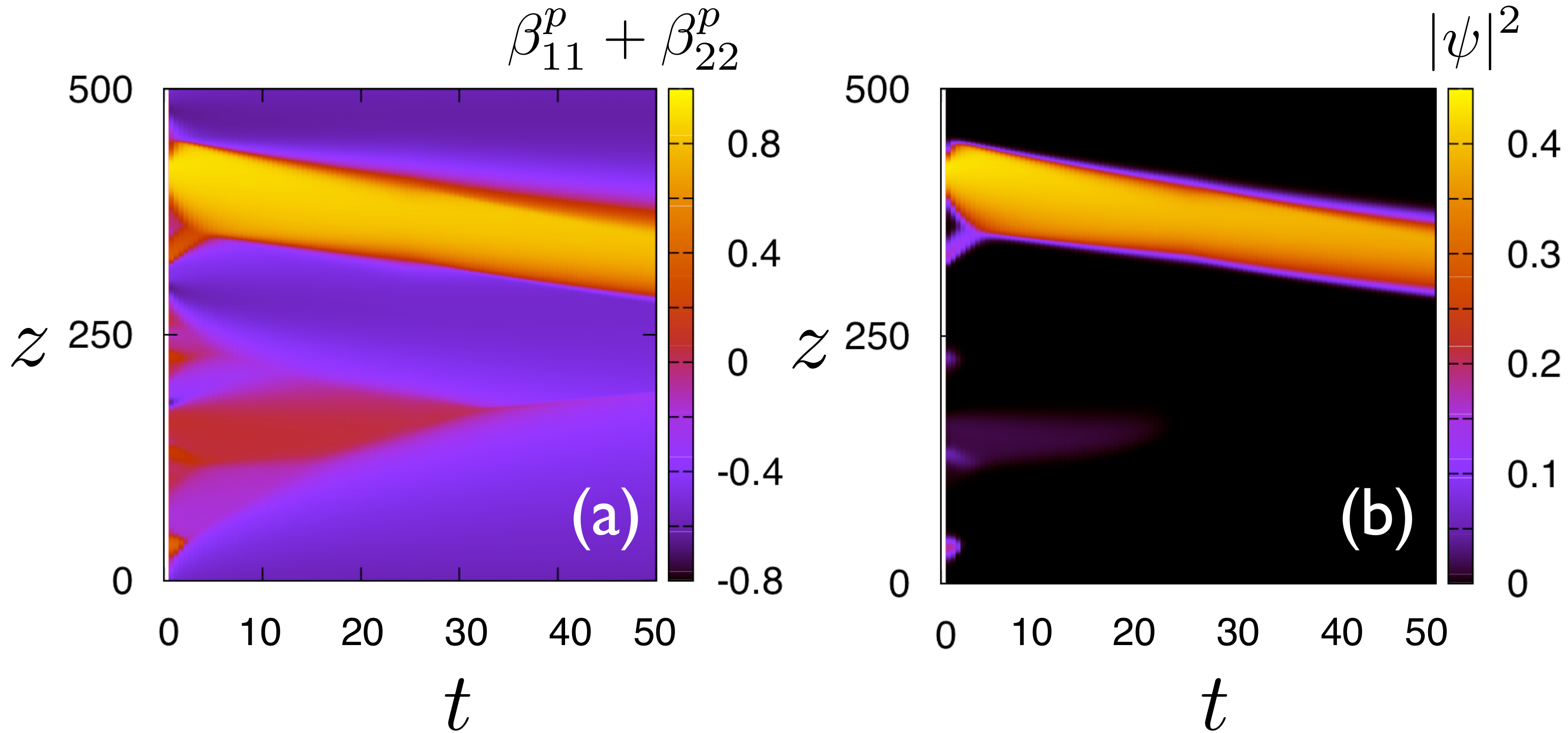}}
\caption{(Color online) Time %otv
evolution for $t\le 50$   %oc 1000\Delta t$ 
of  (a) $\beta^{p}_{11}+\beta^p_{22}\equiv \beta^p_{kk} $ and (b) %ovcd of %otv
 $|\psi|^{2}$ for the 
 system with parameters %oc same 
 as in Fig.\ref{fig:f-and-beta}. The supersolid fraction %ovcd
 becomes large at one of the cell walls where $\beta^p_{kk}$ 
  %otv
is also large (and so is
the absolute value of  the volumetric stress $\sigma_{ii}$). It %ovcd %otv
subsequently decays to zero as the stress relaxes.} %ovcd
\label{fig:time-evolution}
\end{figure}

%otv resordering and some rewording in the next couple of lines
In Fig.~\ref{fig:time-evolution}(a) and (b) we show, respectively, the time evolution of $\beta_{kk}^p$ %ovcd
and of $|\psi|^2$ at these cell walls. 
Initially, there are small pockets of superfluidity at random values of $z$, where the local stress is suitably compressive so that $a < 0$. As a consequence of the plasticity dynamics, dislocations bunch together producing cell walls after some time $t \sim t_0$. Supersolidity is associated {\it only with those cell walls which have a positive stress jump}. While this condition is satisfied at the second cell wall, the first one remains free of $\psi$. Subsequently, 
the stress %oc
decays as $\sim 1/\sqrt{t - t_0}$ \cite{burgulence,Lim-CW} driving the local $a > 0$,
causing $\psi$ to vanish after some time. %oc
%ovcd must be rewritten and so does $|\psi|^2$ since $a \propto \beta^p_{kk}$. Supersolidity at cell walls therefore exists as a long lived transient decaying as $\vert \psi \vert^2 \sim a^{2 \beta} \sim t^{-1/2}$, where $\beta$, the order parameter critical exponent is $1/2$ within our mean field dynamics. 
%The decay of $\psi$ is therefore related to the time taken by the system to anneal out dislocations and produce symmetric grain boundaries which are stress free\cite{gb-super-qmc}. 

%The actual dependence of the NCRIF on annealing time may be further complicated if more than one cell wall is present, where the three dimensional microstructure of the sample becomes important. 
\begin{figure}[h]
\rotatebox{0}{\includegraphics[width=3.5in]{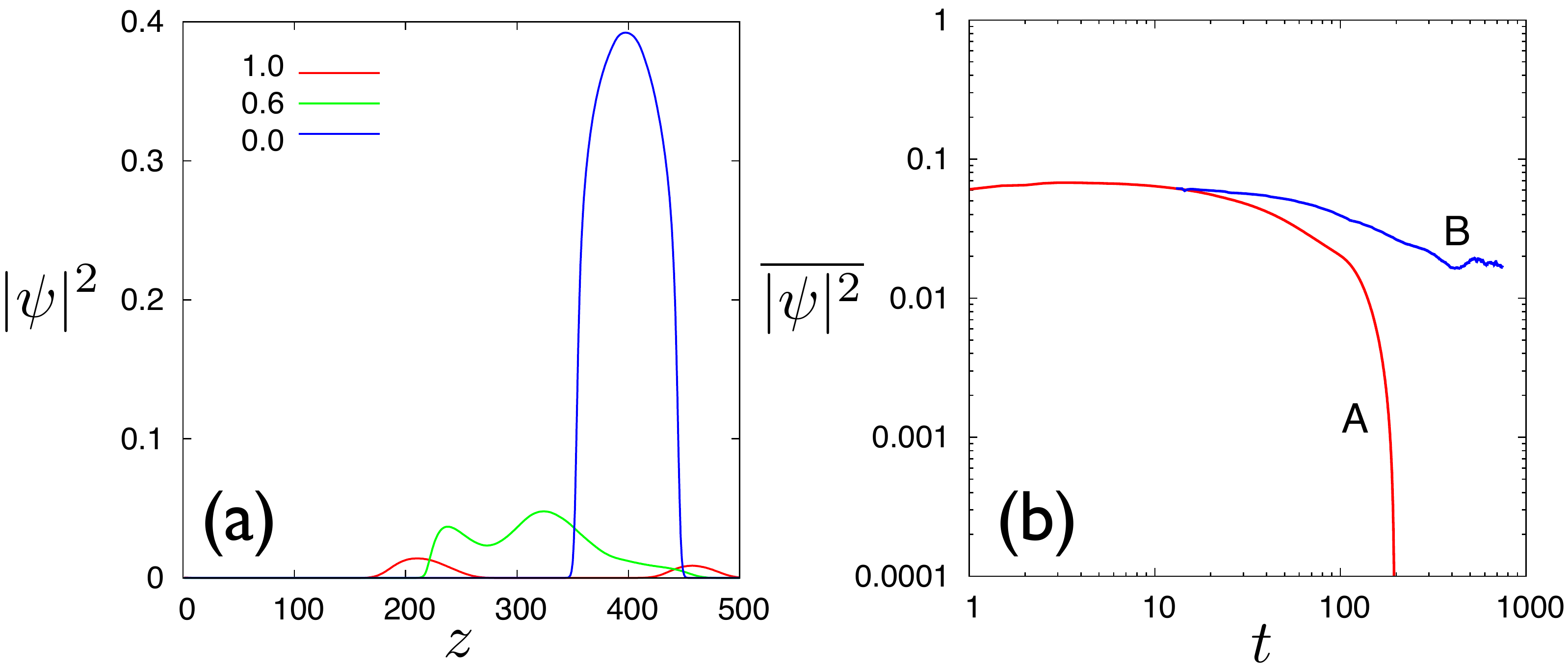}}
\caption{(a) Magnitude of the order parameter at $t = 50$ %oc \Delta t$ 
for $\lambda=1.0$, $\lambda=0.6$ and $\lambda=0.0$. For $\lambda = 1$, formation of regions with large $|\psi|^2$ is suppressed. (b) Spatially averaged $\overline{|\psi|^{2}}$ vs time $t$ for protocols A and B (see text). A:  $\lambda = 0$  %oc
 with other parameters identical to Fig.\ref{fig:time-evolution} and Fig.\ref{fig:f-and-beta}. B:  here %oc %CD deleted spurious )
$\lambda$ was increased to $1$ after $t = 12.5$, note the arrest of the vanishing of  $\overline{|\psi|^2}$ in this case.   %oc
} %oc
\label{fig:couplings}
\end{figure}

%%otv change paragraph break
%The magnitude of the coupling of the elastic degrees of freedom to $\psi$ determines the value of the NCRIF at cell walls. In Fig.~\ref{fig:couplings}(a), the NCRIF is plotted at a particular %otv
%time for various values of $g$. Predictably, larger $g$ enhances NCRIF. The actual values of the full coupling tensor $g_{ij}$ cannot be obtained from a coarse-grained theory such as ours and needs a more microscopic approach or must be measured in careful experiments.  
%%Finally, %otv
%Apart from the superfluid coupling, 
%The nature of the dislocation dynamics also determines supersolid behaviour quite crucially. 
If $\psi$ couples only to the volumetric stress as in our case (and in Ref.~\onlinecite{Toner}), %otv
then without dislocation climb, we do not find %ovcd cannot explain 
supersolid behaviour. This is shown in Fig.~\ref{fig:couplings}(a) %otv
where we have plotted $|\psi|^2$ for different values of $\lambda$. For $\lambda = 1$, when climb motion is completely suppressed (a reasonable limit for solids at low temperatures far from melting), 
supersolidity never appears since $J_{ij}$ is traceless (see Eq. (\ref{current})) and compressive stresses are not spontaneously produced. On the other hand, any pre-existing compressive stress %ovcd, if present, 
relaxes extremely slowly due to the constraints in Eq. (\ref{supersolid}). This suggests a strong link between suppression of dislocation climb and supersolidity \cite{superclimb}. 

In Fig.\ref{fig:couplings}(b) we illustrate this link. %ovcd graphically. 
We show the evolution of the spatially averaged $\overline{|\psi|^2}$ as a function of time %oc
 following two protocols A and B. Protocol A is identical to that shown in Fig.\ref{fig:time-evolution} with $\lambda = 0$. On the other hand in B, we increase $\lambda$ to unity  after simulating with $\lambda = 0$ up to $t=12.5$, %oc %ovcd
 mimicking a decrease in the rate of climb. Fig.\ref{fig:couplings}(b) shows that, in this case, the annealing of $\psi$ is arrested and cells walls with large $|\psi|^2$ may persist at low temperatures to experimentally observable times. Finally, the experimental time over which NCRIF would become %ovcd
unobservable in solid $^4$He does, of course, depends on the parameters $\Gamma$ and $D$. If $\Gamma \geq D$ and the plasticity relaxation time, $D^{-1}$, large, NCRI phenomena will be ubiquitious but may depend sensitively to sample preparation history \cite{sample, anneal}. Further, $D$ is, in general, also dependent on the dislocation density $\rho_{ij}$ (work hardening) and pinning by $^3$He impurities \cite{pinning} at low temperatures, both of which will act to arrest the annealing of $\psi$ and strengthen %oc further ramify 
our conclusions. However, inclusion of such effects may require going beyond the version of continuum plasticity used here \cite{Lim-2,Lim-thesis}. On the other hand if $\Gamma << D$, NCRI phenomena may become altogether unobservable since the required stress relaxes before a non-vanishing $\psi$ can develop. 
% changes below

The results obtained from our coarse-grained description of superfluidity and plasticity are consistent with those of earlier microscopic studies \cite{superclimb,gb-super-qmc} using quantum Monte Carlo methods. In Ref.~\onlinecite{gb-super-qmc}, it was found that grain boundaries separating crystallites that differ from each other by a simple rotation do not support superfluidity. Other, more generic grain boundaries were found to exhibit superfluidity in regions of a few lattice spacings in width. Such microscopic regions of superfluidity will not appear in our coarse-grained description. Supersolid behavior may, of course, arise from a system-spanning network of superfluid channels \cite{Dorsey,Toner,shev} of microscopic width (e.g. cores of dislocation lines \cite{core} and microscopic regions near grain boundaries). A description of this behavior is beyond the scope of our coarse-grained analysis. The importance of dislocation climb in supersolidity, found in the present work, has been emphasized earlier in the microscopic study of Ref.~\onlinecite{superclimb}.

Our work shows that  supersolidity should be observable in solid $^4$He when certain thermodynamic, mechanical and dynamical conditions are satisfied. We find that the degree of supersolid behavior depends crucially on the details of the annealing process. This necessarily implies a large sample to sample variation, as observed in experiments \cite{sample}. Our observation that supersolidity vanishes in the long-time limit if the plasticity field evolves according to its natural dynamics (including dislocation climb) provides a reasonable explanation of the experimental observation \cite{elastic} that the shear modulus of solid $^4$He increases as $T$ is lowered across the onset temperature of the NCRIF. The increase in the shear modulus is generally attributed \cite{elastic} to the pinning of dislocation lines at impurities such as $^3$He atoms \cite{pinning}. Pinning should suppress climb motion, thereby preventing the annealing out of the supersolid field. So, the onset of supersolid behavior and pinning-induced enhancement of the shear modulus should coincide in this picture. Our results on the importance of climb motion in the initial development of the supersolid order
parameter provide an explanation of experimental results indicating that mass flux, dislocation climb and NCRI 
are coupled together in solid $^4$He \cite{mass-flux,core,superclimb}. 
Extensions of this work to investigate the role of $^3$He impurities \cite{pinning} and to obtain the elasto-plastic response of a supersolid with dislocations to external stress and the glass transition \cite{super-glass} in this system are in progress.  
%Finally, we would like to be able to understand phenomena such as glass transition\cite{super-glass} and mass flow in solid $^4$He\cite{mass-flux} which needs a sophisticated understanding of the interplay of plastic degrees of freedom and superfluidity.

The authors acknowledge useful discussions with Sriram Ramaswamy. Financial assistance from a Indo-US Science and Technology  Forum %oc %ovcd
 grant is gratefully acknowledged. 

%otv \begin{references}

%otv \end{references}

\end{document}